\documentclass{aa}
\usepackage{graphicx}
\usepackage{natbib}

\begin{document}
\title{Blazhko stars: always distorted?}

    \author{J. Jurcsik 
    \and J. M. Benk\H{o}
    \and B. Szeidl
          }

%   \offprints{J. Jurcsik}

   \institute{Konkoly Observatory of the Hungarian Academy of Sciences,
              P. O. Box  67, H-1525 Budapest, Hungary\\
              \email{jurcsik,benko,szeidl@konkoly.hu}
             }
   \date{Received; accepted}

   \abstract{Reexamination of the amplitude ratios of the photometric 
and radial velocity changes of Blazhko RRab stars in their different 
Blazhko phases shows that a normal amplitude ratio occurs near the phase 
of maximum amplitude. Checking the Fourier parameters of maximum amplitude 
light curves with different methods proved, however, that none of these 
light curves can be regarded as a normal RR Lyrae type light curve.     
  \keywords{Stars: variables: RR~Lyr --
            Stars: oscillations --
            Techniques: photometric --
            Techniques: radial velocities
            }
   }

   \maketitle
%%%%%%%%%%%%%%%%%%%%%%%%%%%%%%%%%%%%%%%%%%%%%%%%%%%%%%%%%%%%%%%%%%%%%%%%%%%

\section{Introduction}

Long period (10-100 days) amplitude and/or phase modulation 
of RRab stars, often called the Blazhko effect, has been known 
for about a century. Although 30--40\% of the fundamental mode RR 
Lyrae stars exhibit the phenomenon theoretical and observational 
efforts have not led to complete understanding of its physical background.
One crucial point that may help in selecting the correct explanation 
would be to decide what phase of the modulation (if any) corresponds 
to unmodulated single mode RRab radial pulsation i.e. can be regarded 
as normal, undistorted behaviour.

It was pointed out by \citet{SzeidlMM} that on the pulsational   
period -- photometric amplitude diagram the {\it large amplitude}
phases of  Blazhko variables fall within the range occupied by normal 
RRab stars. From the visual inspection of the light curves it can be
seen that peculiar shapes occur in the small amplitude phases. 
These facts indicate that if there is any normal phase of a Blazhko star, 
it should occur during the phases when the amplitudes are larger. 
However, we do not know for sure that such a phase really exists, 
and if it does, whether corresponds to the largest amplitude phase.

In the recent past simple linear relations between the physical properties 
and the Fourier parameters of the light curves of RRab stars were derived 
(\citealt{KW99}, and references therein) which have been proven to give 
similarly accurate results to direct spectroscopic methods. A compatibility 
criterion ($D_m$) of the light curves with those of a large sample of
RRab stars was given in \citet[see also \citealt{JK96}]{KK98}.
This method utilizes the interrelations between the Fourier parameters of 
the light curve and serves as a firmer restriction on the actual values of 
the Fourier parameters than their possible ranges. To quantify the 
morphological compatibility between a given RRab light curve and that
of a large sample of RRab light curves, the $D_m$ parameter was introduced. 
This parameter represents the maximum value of the deviations of the observed 
Fourier amplitudes and phases from their predicted values calculated 
according to the interrelations. Checking whether these relations give
correct values in any phase of the Blazhko cycle can help decide 
which is the normal phase during the Blazhko modulation. 

Another tool for checking similarity (or difference) of a Blazhko star 
to the archetype of the class is to follow their behaviour in the 
{\it photometric amplitude} ($A_V$) 
 -- {\it radial velocity amplitude} ($A_{v_{\rm rad}}$) plane.
This type of diagram was first constructed by \citet{Kur}, who showed that 
these observed quantities are linearly correlated for different types of 
pulsating stars. \citet[hereafter PS67]{PS} documented the behaviour of 
three Blazhko type variables in a similar plot.

In this paper we reexamine the light change variations of Blazhko stars 
utilizing the $A_V$--$A_{v_{\rm rad}}$  relation, compare the measured 
and calculated [Fe/H] values \citep{JK96} at different Blazhko phases, 
and measure the compatibility of the light curves with that of normal 
RRab stars using the interrelations between the Fourier parameters 
\citep{KK98}, in order to answer the question: what is the normal state 
of a Blazhko star?

\section{$A_{v_{\rm rad}}$ \it {vs.} $A_V$}\label{Secav_ap}

\begin{table*}
\caption[]{Compilation of radial velocity and photometric
amplitudes of Blazhko stars}\label{Data}
\begin{tabular}{@{\hspace{-0.1pt}}lcl@{\hspace{-0.1pt}}llclll@{\hspace{-0.5pt}}}
\hline
\noalign{\smallskip}
& \multicolumn{4}{c}{Radial velocity data} &\multicolumn{4}{c}{Photometric data}\\
\noalign{\smallskip}
\cline{2-9}
\noalign{\smallskip}
Name & Time interval$^{\mathrm a}$ & $A_{v_{\rm rad}}^{\mathrm b}$ [km/s] 
&  \multicolumn{1}{c}{$\!\!\!\Phi_{\rm B}$} & Ref.$^{\mathrm {c, d}}$ & Time
interval$^{\mathrm a}$ 
& $\,\,\,A_V^{\mathrm b}$ [mag]&\multicolumn{1}{c}{$\!\!\Phi_{\rm B}$}&$\!\!$Ref.$^{\mathrm c}$\\
\noalign{\smallskip}
\hline
\noalign{\smallskip}
RS Boo &  46951 -- 46959  & $\phantom{>}68\pm1$  & $0.84\pm0.01$ & 2 (M) & 46948 -- 46949  &
$\phantom{>}1.24\pm0.01$ &$0.83\pm0.001$ & 2\\
 &  46879     & $\phantom{>}65\pm1$ &$0.69\pm0.001$& 2 (M) & 38844 -- 38878 &
$\phantom{>}1.13\pm0.02$ &$0.62\pm0.03$ & 1\\
 &     38928 -- 38930  & $\phantom{>}56\pm5$ & $0.75\pm0.002$   &  13 (M)& 43716  &
$\phantom{>}1.18\pm0.02$ & 0.75 & 4 \\
XZ Cyg& 32688 -- 32690 & $\phantom{>}66\pm10^{\mathrm e}$ & $0.0\pm0.02$ &  9 (H+M)& 32689
&$\phantom{>}1.2\pm0.1^{\mathrm f}$ & $0.0\pm0.01$ & 5 \\
      & 32717 -- 32719 & $\phantom{>}48\pm10^{\mathrm e}$ & $0.5\pm0.02$&  9 (H+M)& 33306 -- 33308 &
$\phantom{>}0.8\pm0.1^{\mathrm f}$ &$0.5\pm0.02$ & 5 \\
XZ Dra &  (41078 -- 41168) & $>$67$\pm3$ & $0.81\pm0.07$ & 11 (M)&   (40494 -- 41133)   &
$\phantom{>}0.99\pm0.03$ &$0.83\pm0.02$ & 11,6 \\
       &  (41078 -- 41168) & $\phantom{>}61$:$\pm5$ & $0.18\pm0.09$ & 11 (M)&   (40494 -- 41133)   &
$\phantom{>}0.93\pm0.03$ &$0.18\pm0.12$ & 11,6 \\
       &  38929 -- 38940 &$\phantom{>}43\pm5$ & $0.6\pm0.08$ & 13  (M)& (36410 -- 39403)   &
$\phantom{>}0.72\pm0.05$ &$0.6\pm0.05$ &11 \\
RR Lyr& 37825 & $>$42& 0.81         & 7 (M)&   & $\phantom{>}$0.62:$^{\mathrm g}$& 0.81 & 7 \\
      & 37833  & $>$63& 0.00        & 7 (M)& 37833  & $\phantom{>}$0.965& 0.00 & 7 \\ 
      & 37855  & $\phantom{>}36$&  0.54 & 7 (M)& 37855  & $>$0.5 & 0.54 & 7 \\
      & 37858  & $>$39& 0.61       & 7 (M)    &  & $\phantom{>}$0.5:$^{\mathrm g}$ & 0.61 &7 \\ 
      & 37859  & $>$37& 0.64       & 7 (M) &37859 &    $>$0.465   & 0.64 &7 \\
      & 37863  & $>$37& 0.74       & 7 (M) & 37863  &   $\phantom{>}$0.56    & 0.74 &7 \\
      & 37867  & $>$47& 0.84       & 7 (M) &37867 &   $\phantom{>}$0.71    & 0.84 &7 \\
      & 38215  & $\phantom{>}45$&  0.36           & 7 (M) &38215  &  $\phantom{>}$0.67  & 0.36 &7\\
      & 38219  & $\phantom{>}47$&  0.46           & 7 (M) &38219 &
$\phantom{>}$0.65    & 0.46 &7\\
      & 38240  & $\phantom{>}54$&  0.97           & 7 (M) &38240 & $\phantom{>}$0.79    & 0.97 &7\\
      & 38244  & $\phantom{>}52$&  0.07           & 7 (M) &38244 & $\phantom{>}$0.75    & 0.07&7 \\
      & 38596  & $\ge$47& 0.70            &7 (M)  &38596 &    $\phantom{>}$0.74    & 0.70 & 7\\
      & 38601  & $\ge$40& 0.82            &7 (M)  &38601 &      $>$0.59  & 0.82 &7 \\
      & 43384 -- 43388 & $\phantom{>}45\pm2$  & $0.63\pm0.05$ & 12 (M)&    &
$\phantom{>}$0.70:$^{\mathrm g}$ & $0.63\pm0.05$ &7,10\\ 
RV UMa& (36624 -- 37056) & $\phantom{>}74\pm8^{\mathrm h}$ & $0.07\pm0.07$ & 8 (H)& (36647
--
37061)
&$>$1.38$\pm0.02$ &$0.07\pm0.07$ &8,3 \\
      & (36624 -- 37056)  & $\phantom{>}67\pm10^{\mathrm h}$ & $0.85\pm0.05$  &8 (H)& (36647 --
37061) &
$\phantom{>}1.30\pm0.02$ & $0.900\pm0.005$  &8,3 \\
      & (36624 -- 37056)  & $\phantom{>}40\pm8^{\mathrm h}$ & $0.36\pm0.04$  &8 (H)& (36647 --
37061) &
$\phantom{>}0.72\pm0.02$ & $0.375\pm0.025$  &8,3 \\
\noalign{\smallskip}
\hline
\end{tabular}
\begin{list}{}{}
\item[$^{\mathrm{a}}$]
 JD$-$2\,400\,000. Parentheses mean that observations of the given $\Phi_{\rm_B}$ 
were selected from this interval.
\item[$^{\mathrm{b}}$]
Errors include estimates of systematic discrepancies.
\item[$^{\mathrm{c}}$] 
[1] \citet{Fitch}, [2] \citet{Jones}, [3] \citet{Kanyorv}, [4] \citet{Kanyo}, [5] \citet{Mul},
[6] \citet{Penst} [7] \citet{Pres65}, [8] PS67, [9] \citet{Struve}, [10] \citet{Bela},
[11] \citet{Mitteil}, [12] \citet{Wilson}, [13] \citet{WoAly}.
\item[$^{\mathrm{d}}$]
Lines that the $v_{\mathrm {rad}}$ data were derived
from  (H: Hydrogen, M:Metal) are indicated.
\item[$^{\mathrm{e}}$]
Derived from Van Hoof's measurements, based on using more metallic lines than
Struve's data.
\item[$^{\mathrm{f}}$]
Transformed from white light amplitude; the $A_V$ at the same $\Phi_{\rm_B}$
during the JD 2\,433\,412 -- 2\,434\,309 interval was the same. 
\item[$^{\mathrm{g}}$]
Estimated from the amplitudes during other Blazhko cycles according $\Phi_{\rm_B}$, and the
measured magnitudes of maxima, considering the phase of the 4-year
modulation.
\item[$^{\mathrm{h}}$]
Radial velocities of H lines are transformed into metallic scale according to 
$A^{\mathrm{(Met)}}_{v_{\rm rad}}=0.742 A^{\mathrm{(H)}}_{v_{\rm rad}}$
 (see Fig.~\ref{Figrr}).
\end{list}
\vskip -1mm
\end{table*}
\noindent

Although very few radial velocity observations of Blazhko variables 
have been published since the appearance of the PS67 paper,
the $A_V$ -- $A_{v_{\rm rad}}$ diagram of a more complete 
sample of RR~Lyrae stars given by \citet{Liu} provides the opportunity
and also makes it necessary to refine the positions of  Blazhko
variables on this diagram. In order to do so, the literature was 
searched for radial velocity measurements of known Blazhko variables.

Only radial velocity observations with sufficiently good phase coverage 
to measure the total amplitude could be utilized, provided that a conterporary
photometric V amplitude could also be reliably determined.

Table~\ref{Data} lists  the  radial velocity measurements as well as
the corresponding photometric amplitudes of Blazhko stars. Most of the columns 
are self explanatory. The Blazhko phases ($\Phi_{\rm B}$) of the observations
were calculated using the ephemerides given in Table~\ref{Epoch}. 

\begin{figure}[t!!!]
   \centering
   \includegraphics[width=7.8cm]{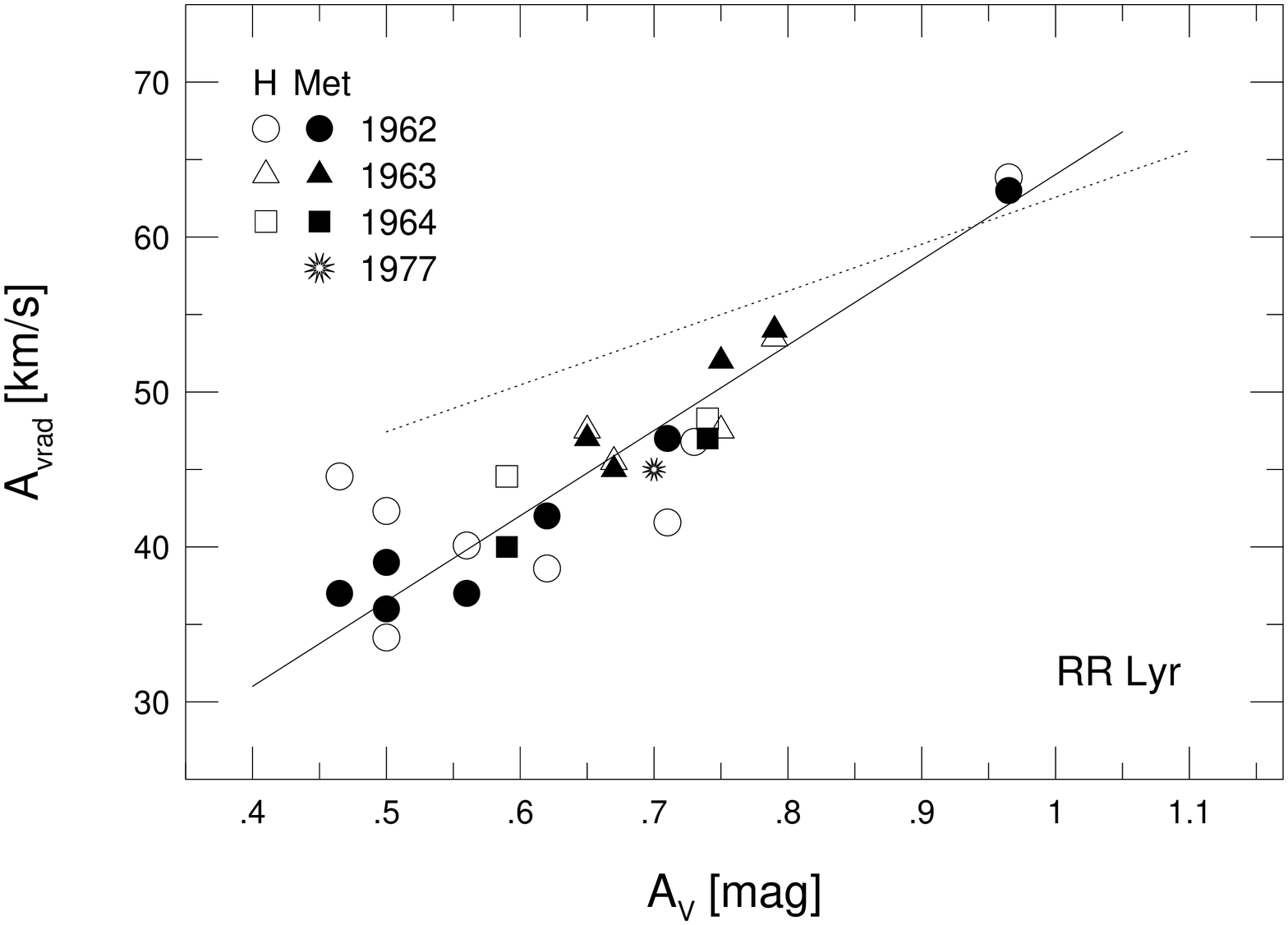}
   \caption{Radial velocity amplitudes derived from metallic and hydrogen 
lines versus photometric $V$ amplitudes of RR~Lyr according to the observations 
of \citet{Pres65} and \citet{Wilson}. Dotted line is for the location of single 
mode RR Lyrae stars \citep{Liu}. The radial velocity amplitudes of H lines
are divided by 1.347 in order to reach agreement with metallic line results. 
The 1963 and 1977 observations correspond to the minimum Blazhko activity phase
of the 4-year cycle of RR~Lyr, whilst the amplitude of the modulation
varies within the 0.1--0.4 mag range.}
              \label{Figrr}
    \end{figure}

The radial velocity data are very heterogeneous, obtained by using instruments 
of different resolutions. A further inconsistency arises from the selection of 
spectral lines used in determining the radial velocity. The relation between the 
radial velocity and photometric amplitudes of pulsation defined by \citet{Liu} 
was based on radial velocities of metallic lines, so we considered metallic line 
velocities if they were given separately from the results for hydrogen lines.   
In the case of RV~UMa, however, only radial velocities of H lines were 
published in PS67. We attempted to transform these to match the metallic line
radial velocities utilizing the extended observations of RR~Lyr \citep{Pres65}. 
The Balmer lines originate from higher regions of the atmosphere exhibiting 
a larger velocity gradient than shown by the photospheric metallic lines.
It was found that radial velocities measured from H and metallic lines  are 
strongly correlated, and with an appropriate linear transformation can be 
equalized within their error ranges (see Fig.~\ref{Figrr}). Although this 
relation may differ from star to star, we transformed the PS67 radial 
velocities of RV~UMa accordingly, to estimate its metallic line radial velocity
amplitudes in different Blazhko phases.

\begin{table}[!ht]
\caption[]{Blazhko modulation ephemerides  }\label{Epoch}
\begin{tabular}{@{\hspace{-0.001pt}}ll@{\hspace{-0.001pt}}l}
\hline
\noalign{\smallskip}
 Name      &  Max Epoch [JD$-$2400000] & ${\rm P}_{\rm B}$ [d] \\
\noalign{\smallskip}
\hline
 \noalign{\smallskip}
  RS Boo &  42786.0   &  532.0   \\
  XZ Cyg &  17022.0 &  \,\,\,57.39   \\
  XZ Dra & 36463.5    & \,\,\,75.8       \\
         & 40707.5    & \,\,\,76.6      \\
  AR Her &  49959.5 &  \,\,\,31.85      \\
  RR Lyr &  34565;\,38665;\,41544;\,42622;\,43401;\,44147   &    \,\,\,40.80   \\
  RV UMa &  36733.7;\,37815.4   &  \,\,\,90.785   \\
\noalign{\smallskip}
\hline
\end{tabular}
\begin{list}{}{}
\item
Ephemeris of XZ~Cyg is taken from \citet{Smith}.
Blazhko periods and maximum epochs for the other stars %\begin{tabular}{lll}
are determined from the collections of photometric data by \citet{Bela}, 
\citet{Mitteil}, \citet{Andrea},  and \citet{Geza} for
RR~Lyr, XZ~Dra, RS~Boo, and RV~UMa, respectively.
Different ephemerides for  RR~Lyr and XZ~Dra
take into account the phase shifts, and period changes
of the modulation.
\end{list}

\end{table}

If there was no photometry simultaneous with the radial velocity observations, 
but the photometric amplitude for the same Blazhko phase could be reliably
estimated from close (or distant) photometric observations,
Table~\ref{Data} lists these amplitudes.  

\citet{Liu} used the $A_V$ -- $A_{v_{\mathrm {p}}}$ relation, where 
$A_{v_{\mathrm {p}}}$ is the pulsational velocity amplitude.
Since the transformation of $A_{v_{\mathrm {rad}}}$ into $A_{v_{\mathrm {p}}}$
involves further uncertainties, we decided to use the directly measured radial
velocity amplitudes. The regression of the $A_V$ -- $A_{v_{\mathrm {rad}}}$  
data of the sample used by \citet[][and references therein]{Liu}  has similar 
error statistics as the $A_V$ -- $A_{v_{\mathrm {p}}}$  dataset, justifying our
choice.

\begin{figure}
   \centering
   \includegraphics[width=8.8cm]{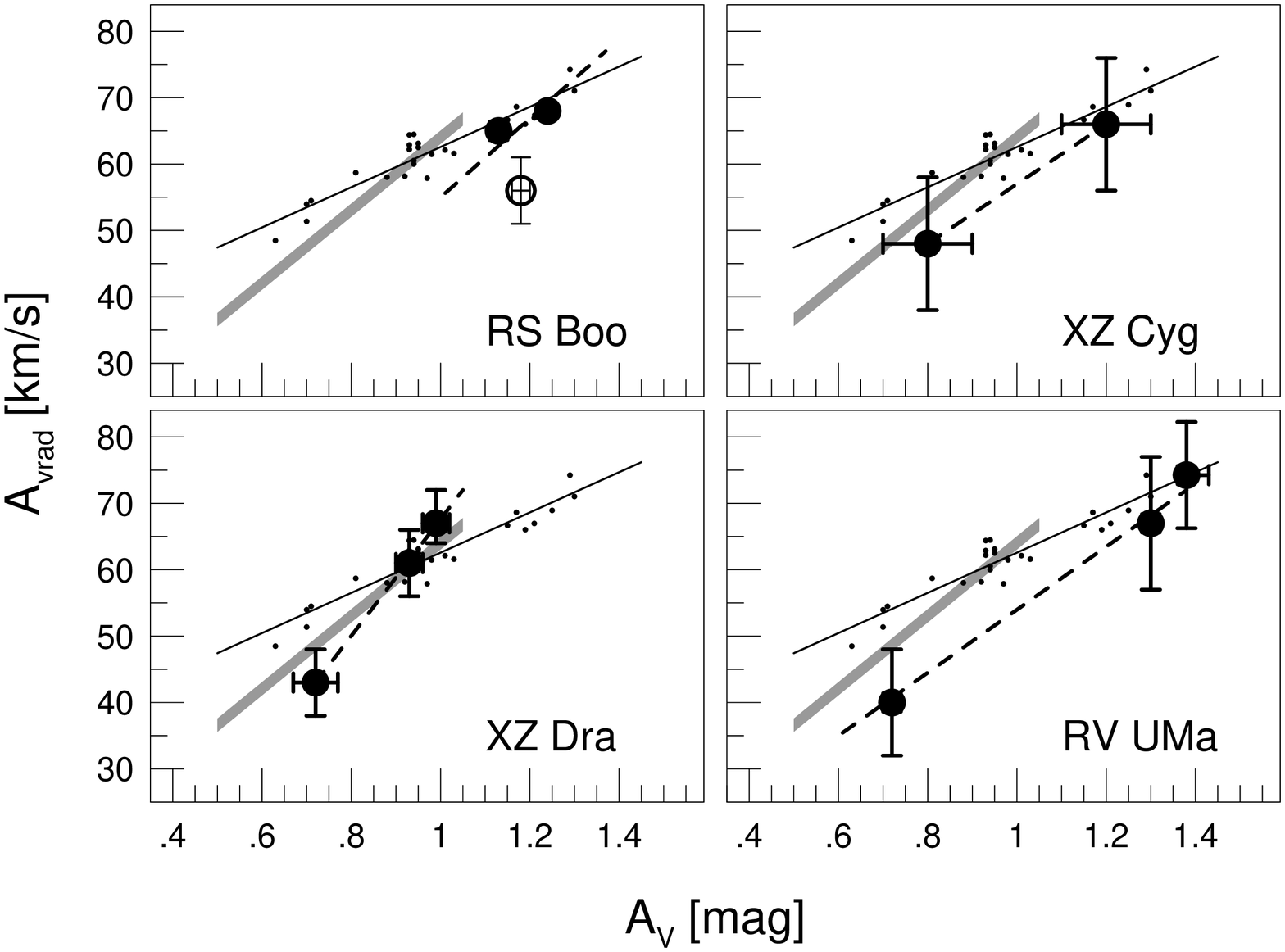}
   \caption{Comparison of the radial velocity and photometric amplitudes
of Blazhko type RR Lyrae stars with the sample of non Blazhko RRab
stars (solid line fitted to small dots using the compilation of \citealt{Liu}).
The greyscale strip, shown in detail in Fig.~\ref{Figrr},
 denotes the positions of RR~Lyr. 
Blazhko stars lie along nearly parallel lines that 
match the area of normal RRab stars in the larger amplitude 
phases. }
              \label{Figav_ap}
    \end{figure}

Fig.~\ref{Figav_ap} shows the behaviour of  5 Blazhko variables in the
$A_V$ -- $A_{v_{\rm rad}}$ plane in different phases of their Blazhko cycles.
In contrast to the early suggestion of PS67 the Blazhko stars seem to behave 
very similarly. 

{\it a)} In the $A_V$ -- $A_{v_{\rm rad}}$ plane they follow tracks with 
the same gradient within the limits of the observational uncertainties, 
these are steeper than the $A_V$ -- $A_{v_{\rm rad}}$ relation defined by 
non-Blazhko variables. This means that the {\it photometric amplitude of 
Blazhko stars in their smaller amplitude phase is  larger than that of a 
normal RRab star with the same radial velocity amplitude.}

{\it b)} The amplitude ratio of their photometric and radial velocity
changes corresponds to that of non-Blazhko stars within the $0.1-0.2$~mag 
vicinity of their maximum photometric amplitudes. 

   \begin{figure*}
   \includegraphics[width=18cm]{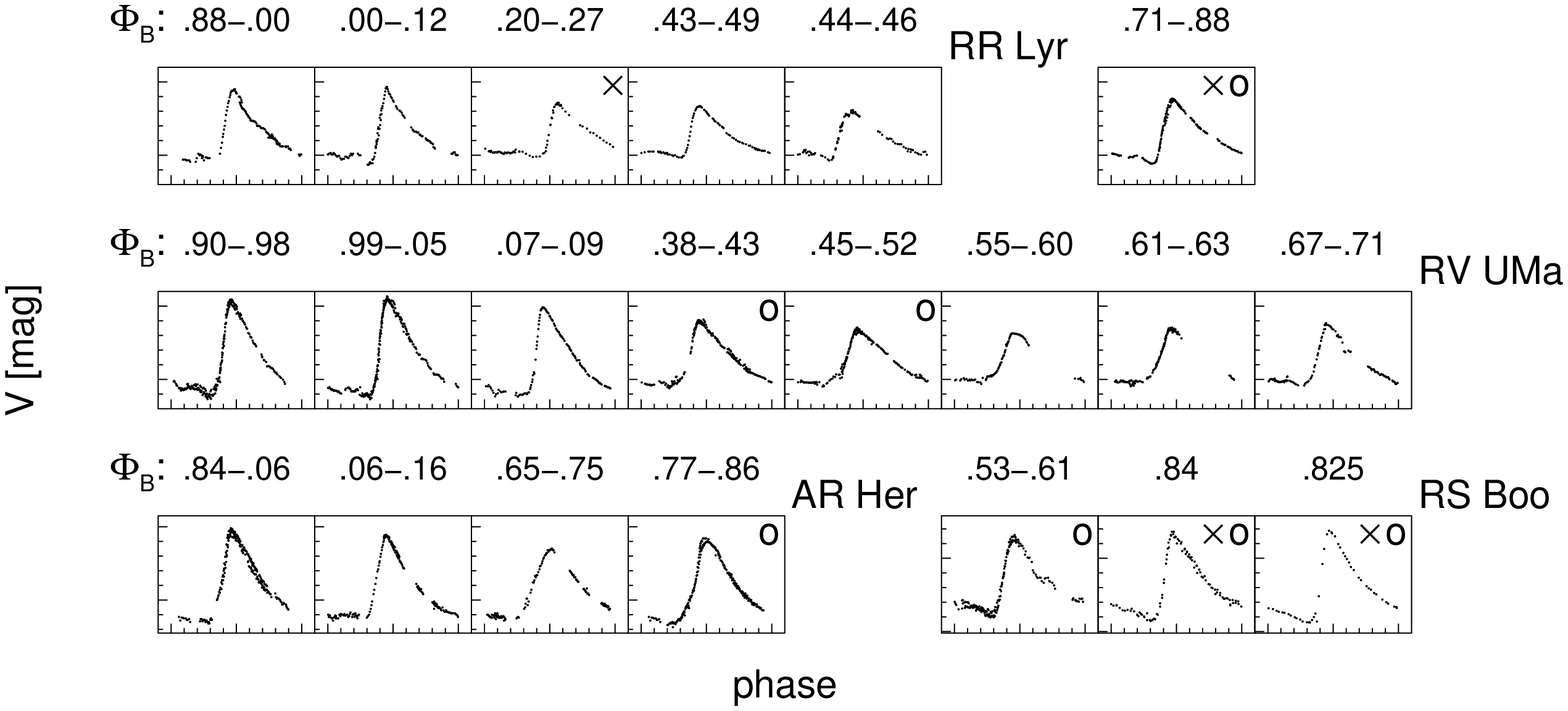}
   \caption{Complete $V$ light curves of Blazhko variables during different
phases of the modulation. Spacing of tick marks are 0.1 phase and 0.2 mag,
respectively.
Blazhko phases according to Tab.~\ref{Epoch} are indicated in the top of the panels.
For RR~Lyr, data of \citet{Murn77, Murn81}, \citet{Fitch}, \citet{Siegel} and \citet{Hardie} 
are used, observations of AR Her are taken from \citet{Smith99}, and the compilations
of photoelectric observations of RV~UMa and RS~Boo \citep{Geza, Andrea} are utilized.
Crosses (x) and circles (o) in the top right corners indicate
$D_m<2$ and $\vert\Delta\rm{[Fe/H]}\vert<0.15$, respectively.}
              \label{Figlc}
    \end{figure*}

\section{Light curve shapes}

The results shown in Sect.~\ref{Secav_ap} hint that  normal behaviour phases 
of Blazhko stars, if any, should be in the larger amplitude phase. 
To check the compatibility of Blazhko stars' light curves with that of 
unmodulated single mode RRab stars we utilize the Fourier decompositions of 
complete light curves of different Blazhko phases. The compatibility is verified 
if both the $D_m$ criterion according to the formalism of~\citet{KK98} is valid, 
and [Fe/H] calculated from the light curve parameters using the \citet{JK96} 
metallicity formula is in good agreement with direct spectroscopic measurements.
Selection criteria are chosen at the 2$\sigma$ limit for $D_m$ and 
$\pm0.15$ dex for the difference between the measured and calculated 
metallicity values ($D_m<2$; $\vert\Delta\rm{[Fe/H]}\vert<0.15$).

The observational strategy followed by observers of Blazhko type stars has
concentrated mostly on the rising branch and the maxima, thus only very 
limited complete light curves of Blazhko stars are available. We have 
constructed 21 nearly complete light curves of four Blazhko variables
with spectroscopically known metallicities. These light curves, shown in 
Fig.~\ref{Figlc} are compiled from the observations of consecutive nights 
of the same Blazhko phases within a limited time interval, in order to 
evaluate the occurrence of long term changes of any type. 

In Fig.~\ref{Figlc} light curves with $D_m<2$, and 
$\vert\Delta\rm{[Fe/H]}\vert<0.15$  are denoted by circles and crosses 
in the top right corners, respectively. Only three light curves fulfil 
both requirements:  two light curves of RS~Boo and one of RR~Lyr.
This light curve of RR~Lyr is however, not representative for the large 
amplitude modulation, as these observations were obtained during the 
minima of the 4-year cycle (JD 2\,434\,553 --2\,434\,560), when the
amplitude of the modulation was smaller than 0.1 mag \citep{DSz}.
The behaviour of RS Boo differs from other Blazhko type variables in 
many respects, it has an unusually short fundamental mode period, a very long 
modulation period and only reduced changes in its light curve shape.
It is important to note that none of the large amplitude light curves 
of RV UMa fulfil both criteria, although from Fig.~\ref{Figav_ap}  
it could be suspected that the large amplitude phase might be undistorted.

\section{Conclusions}

A combination of the $A_V$ -- $A_{v_{\rm rad}}$ behaviour and the 
light curve shapes of Blazhko variables leads to the conclusion that 
for Blazhko stars with large amplitude modulation there is no Blazhko 
phase which is compatible with that of a non-Blazhko RRab star. 
Although the amplitude ratio of the radial velocity and the light change
match the normal mode pulsation during larger amplitude phases,
none of these light curves has been proven to be undistorted.

Stars which seem to agree  with single mode RRab stars in all the studied 
aspects are {\it RR~Lyrae in its low modulation amplitude phase of the 4-year 
cycle}, and RS~Boo whose amplitude modulation range is only about 0.15 mag.

This result means restriction for the theoretical explanation of the
Blazhko phenomenon. Explanations that predict an undistorted light curve
along any phase of a large amplitude modulation seem to be unrealistic.

\begin{acknowledgements}
We thank G\'eza Kov\'acs and Andrea Nagy for providing us with the electronic   
compilations of RV~UMa and RS~Boo data, and to Andrew Wilkins for 
correcting the language of the manuscript. This research has made use 
of the SIMBAD database, operated at CDS Strasbourg, France. 
This work has been partly supported by OTKA grants T30954 and T30955.
\end{acknowledgements}

\bibliographystyle{aa}

\end{document}